\documentclass[final,3p,times,twocolumn]{elsarticle}

\usepackage{graphics}

\usepackage{amssymb}

\journal{Journal of Luminescence}

\begin{document}

\begin{frontmatter}

\title{Light storage protocols in Tm:YAG}

\author{T. Chaneli\`ere \corref{cor1}}
\ead{thierry.chaneliere@lac.u-psud.fr}
\cortext[cor1]{Corresponding author}
\author{R. Lauro}
\author{J. Ruggiero}
\author{J.-L. Le~Gou\"et}
\address{Laboratoire Aim\'e Cotton, CNRS-UPR 3321, Univ. Paris-Sud, B\^at. 505, 91405 Orsay cedex, France}

\begin{abstract}
We present two quantum memory protocols for solids: A stopped light approach based on spectral hole burning and the storage in an atomic frequency comb. These procedures are well adapted to the rare-earth ion doped crystals. We carefully clarify the critical steps of both. On one side, we show that the slowing-down due to hole-burning is sufficient to produce a complete mapping of field into the atomic system. On the other side, we explain the storage and retrieval mechanism of the Atomic Frequency Comb protocol. This two important stages are implemented experimentally in Tm$^{3+}$- doped yttrium-aluminum-garnet crystal.
\end{abstract}


\end{frontmatter}


\section{Introduction}\label{intro}
The quest for a quantum memory for light initially involved atomic gases since the particles in a vapor are well isolated from the environment. This approach led to important experimental realizations in the past few years \cite{julsgaard2004edq, chaneliere2005sar, eisaman2005eit}. The advantages of doped solids have rapidly drawn attention. Being motionless, they can exhibit very long coherence time \cite{longdell2003sls}. The Rare-Earth Ion doped Crystals (REIC) have remarkable properties in that prospect and are widely available. Their feature are also original as compared to atomic gases.

Even if the material properties are interesting, a quantum storage protocol has still to accommodate two conflicting stages. The signal should be stored and then strongly interact with the system. It should be retrieved afterwards without any perturbation to preserve the fidelity of the memory. This fundamental observation usually renders the protocol counter-intuitive at first sight. This is usually an interplay between absorption and transparency of the medium. The former is desired for a good mapping of the field into the atomic excitation. The latter is appreciated at the retrieval stage to avoid perturbations. The protocols that we present here involve both, absorption and transparency, and are exemplary in that sense. Our motivation is also to properly explain this point.

The existence of shelving state with a long population lifetime makes possible the observation of spectral-hole burning. This technique is a common ingredient for the two protocols that we present. Since the population is not shuffled by collisions as in vapors, the hole lifetime time can be very long (up to hours). This method has been a powerful high-resolution spectroscopic tool \cite{volker1989hole}. More generally, it permits an initial spectral tailoring of the inhomogeneous broadening.

The optical coherence lifetime of rare-earth ion can also be long at low temperature (microsecond to millisecond range) thus allowing coherent driving of the electronic transition. In terms of light storage, a direct mapping of the field into the optical coherences gives a significant memory effect at least as a primary stage. The information can then be coherently transfered to long-lived states (nuclear spin for example). The Atomic Frequency Comb protocol (AFC) precisely involves these two successive steps. Inspired by the photon-echo technique, the signal is first absorbed by a spectral periodic structure (the frequency comb). This structure can directly produce an echo. Under appropriate conditions that we analyze in this paper, the signal is efficiently retrieved. We pay particular attention to the efficiency of the recovery process. The echo has been demonstrated to be faithful in principle \cite{AFCTh, AFCexp} making this protocol valuable for a quantum memory. A long storage time is achieved by performing a Raman transfer into the spin coherences. This permits an on-demand retrieval. This is also critical element to accomplish a complete retrieval as we will see in section \ref{AFC}. The Spectral hole burning for Stopping Light (SHBSL) is very different at first sight. The signal is not directly absorbed since the medium shows a transparency target window. The primary interaction is then mainly due to dispersion and gives rise to slow light propagation. We will precisely explain in section \ref{SHBSL} how this effect already involves a mapping of the field into the atomic medium. For a complete light-stopping, a Raman transfer is also performed as in the AFC. This conversion is reversed to trigger the retrieval.

We use the same formalism to describe both protocols.

The experimental realizations for both are implemented in a Tm:YAG crystal. As other non-Kramer's ions, thulium can have a long coherence time without magnetic field, especially in YAG ($\sim$ 100 $\mu$s for low concentration samples). Under magnetic field, the ground and excited singlets split into Zeeman doublets. This structure is relevant to tailor a simple $\Lambda$-system \cite{de2006experimental}, a common ingredient for many coherent manipulations. The ground state splitting can be extensively adjusted by varying the magnetic field intensity (up to 400MHz/T). The interaction wavelength ($\sim$ 793 nm) is also particularly convenient for excitation (laser diode) and detection (high efficiency silicon photo-diode).

\section{Spectral hole burning for stopping light} \label{SHBSL}
The SHBSL initially demands the preparation of a spectral hole in the absorption band. This stage is usually possible in REIC by pumping to a population shelving state. The resulting Transparency Window (TW) is our interaction bandwidth. One expects a very small interaction between the field and the medium. This is actually not true. Although counter-intuitive, the propagation is associated with a complete conversion of the field into the atomic coherences. This effect is due to the slow-light propagation in the TW. At the retrieval stage, the transparency of the medium will be an advantage since the signal can leave the medium without re-absorption. Even if the slow-light propagation is a crucial step of the protocol, it also involves a Raman transfer that only ensure the complete trapping of the light. The SHBSL has some obvious similarities with the Electromagnetically Induced Transparency (EIT) and the stopped-light experiments that have been realized in atomic vapors \cite{phillips2001sla}. We finally present the experimental results obtained in a Tm:YAG crystal.

\subsection{Slow-light propagation in a Transparency Window}
\begin{figure}[pth]
\includegraphics[width=7cm]{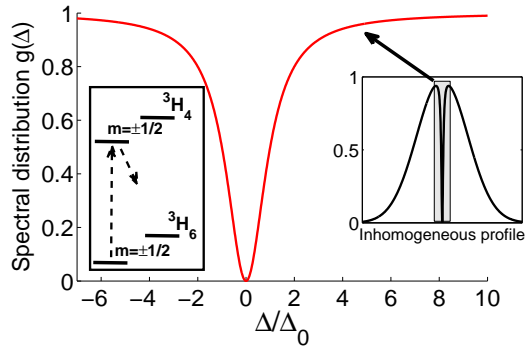}\caption{Within the inhomogeneous profile (right inset), a narrow hole is burnt (width $\Delta_0$). Left inset: schematic thulium level structure (see section \ref{SHBexp} for details).}
\label{fig:Fig_g}%
\end{figure}

We first describe the propagation of a weak field in a TW. This interaction bandwidth is prepared by spectral hole-burning, a well-know optical pumping mechanism in REIC. A narrow hole is tailored within the absorption profile that is described by the spectral distribution $g\left(\Delta\right)$ (see fig. \ref{fig:Fig_g}).

We assume that the pulse bandwidth is much smaller that the hole width $\Delta_0$. The medium is then transparent. We use the Bloch-Maxwell equations assuming the slowly varying amplitude and the rotating wave approximations \cite{allen1987ora}. In the weak signal limit \cite{crisp1970psa}, the atomic population $\mathrm{w}(\Delta;z,t)$ is unchanged and is equal to $-1$ to the first order:

\begin{eqnarray}
\partial_z\Omega(z,t)+ \frac{1}{c}\partial_t\Omega(z,t)=
-\displaystyle\frac{i \alpha}{2\pi}\int_\Delta g\left(\Delta\right) \mathcal{P}(\Delta;z,t) \label{MB_M}\\
\partial_t\mathcal{P}(\Delta;z,t)=-\left(i \Delta + \gamma\right)\mathcal{P}(\Delta;z,t)
-i\Omega(z,t)\label{MB_B}
\end{eqnarray}

where $\Omega(z,t)$ is the Rabi frequency of the field and $\mathcal{P}(\Delta;z,t)=\mathrm{u}(\Delta;z,t)+i\mathrm{v}(\Delta;z,t)$ is the polarization including the in-phase and out-of-phase components of the Bloch vector. The absorption coefficient is $\alpha$, $\gamma$ is the homogeneous line-width.

Under realistic conditions, the hole is usually larger than the homogeneous broadening. This is particularly true when the medium is very absorbing. The hole is indeed broadened by saturation of the pumping process. If we then assume that the signal bandwidth is much smaller than the hole-width, one can describe the polarization by a first order expansion (integral solution of eq. \ref{MB_B}). 
\begin{equation}\label{TF_Bloch}
{\mathcal{P}}\left(\Delta;z,t\right)=-\frac{{\Omega}\left(z,t\right)}{\Delta}-i\frac{{\partial_t\Omega}\left(z,t\right)}{\Delta^2}
\end{equation}
This situation precisely corresponds to the adiabatic following problem in the no-damping limit \cite{allen1987ora}. The atomic evolution is mainly driven by the in-phase component $\mathrm{u}(\Delta;z,t)=-\Omega(z,t)/\Delta$. The out-of-phase component $\mathrm{v}(\Delta;z,t)=-\partial_t\Omega(z,t)/\Delta^2$ is much smaller than $\mathrm{u}$ as pointed out by Grischkowsky \cite{grischkowsky1973afa} in the context of slow-light.

To solve the propagation equation, we restrict the integration over the spectral distribution $g\left(\Delta\right)$ to the vicinity of the hole (width $\Delta_0$). It is assumed to have locally a lorentzian profile $g\left(\Delta\right)\simeq1-\displaystyle \frac{1}{1+4\Delta^2/\Delta_0^2}$. After integration, the in-phase components will vanish, only remains the terms proportional to $\partial_t\Omega(z,t)$:

\begin{equation}\label{Prop_Vg}
\partial_z\Omega(z,t)+ \frac{1}{c}\partial_t\Omega(z,t)=-\frac{\alpha}{\Delta_0}\partial_t\Omega(z,t)
\end{equation}
In other words, the propagation is described by the group velocity $V_g$ defined as
\begin{equation}\label{Vg}
\frac{1}{V_g}=\frac{1}{c}+\frac{\alpha}{\Delta_0}
\end{equation}

The pulse propagation is then essentially determined by a dispersive effect. Under appropriate conditions, the signal is not absorbed but simply delayed by passing through the medium. In the extreme situation of slow-light propagation $V_g << c$ requiring a strong absorption $\alpha$ and narrow hole, the pulse is strongly spatially compressed and can be entirely contained inside the medium. That's the ideal situation that we consider for our protocol. A simple way to understand this interaction is to consider the energy carried by the propagating field. It is given by a spatial integration of the intensity proportional to $\displaystyle \int_z \Omega^2(t-z/V_g)$. The direct consequence of the pulse spatial compression by a factor $V_g/c$ is the energy reduction by the same amount. As soon as $V_g << c$, the field contains a very small fraction of the incoming energy which has been fully transfered to the atomic system. If the field energy is given by $U_{\Omega}=\int_z \Omega^2(z,t)$, one can derive from the Bloch-Maxwell equation the energy stored in the atomic population. With the same units, it is given by $U_{w}=\displaystyle \frac{\alpha c}{\pi} \int_z \int_\Delta \mathrm{w}(\Delta;z,t) g\left(\Delta\right)$. To the second order, the population deviates from $-1$ because of the small contribution of in-phase component: $\mathrm{w}=-1+\mathrm{u}^2/2=\displaystyle -1+\frac{\Omega^2}{2 \Delta^2}$. One can then verify that inside the medium $U_{\Omega}$ is only a $V_g/c$ fraction of the incoming energy. The rest $1-V_g/c$ is indeed stored into the atomic system $U_{w}$.

In the prospect of light storage, this is an interesting situation since the field has been entirely mapped into the atomic variables. Nevertheless, this is not sufficient to ensure a complete capture of the field information.

\subsection{From field mapping to complete storage}\label{mappingstorage}

As we have seen, the slow-light propagation due to strong dispersion is also associated with a complete energy transfer to the atoms. Even if the field only contains a small energy fraction, it still controls the propagation. The atomic variables are adiabatically following the field. In that sense, the field mediates the propagation of the atomic excitation at the group velocity. This may sound surprising since the field energy is very weak. But it is not contradictory because total flux through a surface perpendicular to the propagation is obviously conserved as pointed by Courtens \cite{PhysRevLett.21.3} in the context of Self-Induced Transparency. Without external control, the energy is transfered back to the field at the end of the medium leaving the atomic system in its initial state.

\begin{figure}[pth]
\includegraphics[width=7.5cm]{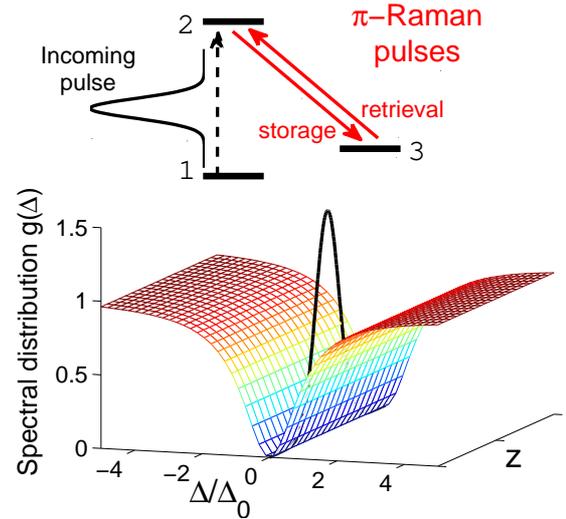}\caption{During the slow-light propagation, if the optical thickness is high, the incoming pulse on the 1-2 transition is fully contained inside the hole-burnt medium (spectral distribution $g\left(\Delta\right)$). The storage and retrieval are controlled by $\pi$ Raman pulses (see text for details).}
\label{fig:FigSchemeSHBSL}
\end{figure}

To freeze the atomic evolution and store the information, we propose to convert the optical dipoles into Raman coherences. A short $\pi$-pulse is applied on the 2-3 transition (see fig. \ref{fig:FigSchemeSHBSL}). If the Raman transfer is sufficiently fast, the optical dipoles vanish. As a consequence, the group velocity equals $c$ and the remaining field leaves the medium rapidly. Without field mediating the propagation, the atomic excitation stays in place and is stored in the Raman coherences 1-3. The conditions that the Raman pulses should fulfill are realistic and have been carefully analyzed \cite{LauroTh}. The storage time is then limited by the lifetime of the Raman coherences. It can be long for ground state hyperfine levels in REIC.

The retrieval is also triggered by a Raman transfer on the 2-3 transition (second $\pi$-pulse). This back-conversion restores the optical coherences that have been initially excited by the signal. These dipoles radiate in the TW and produces a field that can leave without being reabsorbed. For a very strong absorption $\alpha$ ensuring that the signal is entirely compressed inside the medium, the retrieved pulse is a faithful copy of the incoming signal \cite{LauroTh}. This constraint on the optical thickness is a common feature of the SHBSL and the EIT based protocols as we will see in section \ref{EITcomp}.
 
\subsection{Slow-light propagation in Tm:YAG}\label{SHBexp}

The SHBSL protocol is experimentally investigated in a 0.5\% Tm-doped YAG crystal. The ground state $^{3}\textrm{H}_{6}$ and the excited state $^{3}\textrm{H}_{4}$ split into two Zeeman sublevel under magnetic field (see fig. \ref{fig:Fig_g} left inset). This permit an efficient hole-burning process since the population lifetime in the ground state is long at low temperature (typically few seconds below 2K). The orientation of the crystal in the magnetic field has been chosen to optimize the branching ratio between the Zeeman sublevels as decribed in ref \cite{louchet:035131}. The homogeneous width is $\gamma\simeq10$ kHz. To prevent any broadening from the laser jitter, our extended cavity diode laser operating at 793 nm is stabilized on a high finesse Fabry-Perot cavity through a Pound-Drever-Hall servoloop. Narrow structures can then be tailored within the inhomogeneous line ($\simeq$10GHz).

\begin{figure}[pth]
\includegraphics[width=7cm]{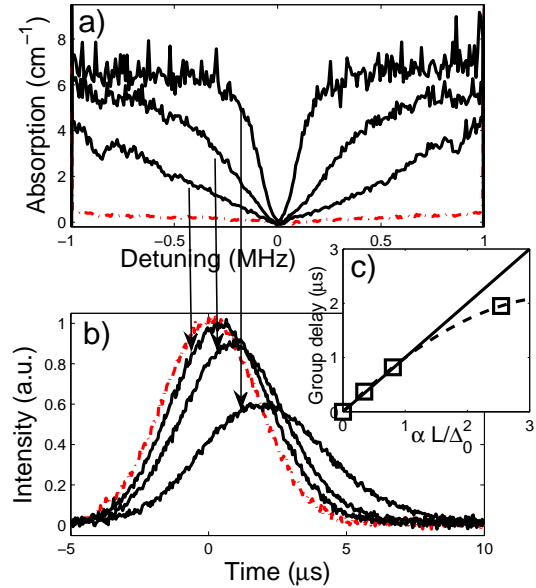}\caption{a). Transmission spectrum for different hole-burning power. b). The brief probe pulse (rms width 1.75$\mu s$) is delayed depending of the hole width. c). The corresponding group delays at compared with the expected ones $\frac{\alpha L}{\Delta_0}$.}
\label{fig:TraitSLHB}
\end{figure}

A long monochromatic pulse (200ms) is first applied to burn a hole. A repumping procedure is also used to increase the absorption coefficient up to 7 cm$^{-1}$ \cite{LauroExp}. The width of the hole strongly depends on the power of the pumping beam. This power broadening effect is imposed by the population dynamics. It allows us to change $\Delta_0$ and then to obtain different group velocities. In order to compare the expected group delays with the measurements (eq. \ref{Vg}), we first record the transmission spectrum to measure independently the width of spectral distribution (see fig. \ref{fig:TraitSLHB}a). We then send a brief probe pulse (rms width 1.75$\mu s$) through the sample and measure the delay for different hole width (see fig. \ref{fig:TraitSLHB}b). To obtain a reference pulse, we first burn a 10MHz pit (red dashed line in  fig. \ref{fig:TraitSLHB}a). The corresponding transmitted pulse is neither delayed nor absorbed (red dashed line in fig. \ref{fig:TraitSLHB}b). 

We clearly observe that the group delay strongly depends on the hole width (see fig. \ref{fig:TraitSLHB}b). It equals $\displaystyle \frac{\alpha L}{\Delta_0}$ (eq. \ref{Vg}) in the slow-light case. Since we can extract this experimental parameters from the spectral distribution (fig. \ref{fig:TraitSLHB}a), we compare the expected and measured delays for different hole width (fig. \ref{fig:TraitSLHB}c). The agreement is good for broad holes (short delays). A discrepancy appears for the narrowest hole. In that case, as we can see in fig \ref{fig:TraitSLHB}c, the pulse is significantly absorbed (intensity reduction) and deformed. Our simple propagation theory (eq. \ref{Prop_Vg}) is not valid anymore. An higher order expansion of the spectral distribution would be required to account for the pulse absorption and distortion. This explains the difference with the expected value of the group delay.

To store the information by stopping light, one would have now to apply a Raman pulse as explained in section \ref{mappingstorage}. The long coherence time of the Zeeman states and the branching ratio would be appropriate for this operation \cite{louchet:035131}. Nevertheless this is not possible since we already use this level as a population shelving state (see fig. \ref{fig:Fig_g} left inset). Such a demonstration would be possible with REIC with an hyperfine structure at zero magnetic field then benefiting from a manifold in the ground state. Regarding thulium, one can also use the Zeeman doublet of the excited state where population can be kept for few hundreds of microseconds. This should be sufficient to burn a hole by shelving the population in the excited state then leaving the ground state doublet for the Raman transfer. This possibility is currently under investigation in our group.

\subsection{Comparison with EIT based protocol}\label{EITcomp}

The SHBSL has some obvious similarities with the stopped-light storage produced by EIT. The spatial compression is a critical ingredient since the pulse should fit inside the medium \cite{lukin2003cta}. It is due to the steep dispersion profile whatever is the physical origin of the TW. As a consequence, the requirements relating the interaction bandwidth and the optical thickness are the same as in EIT \cite{PhysRevLett.84.5094}. If $L$ is the length of the medium, $\sqrt{\alpha L}$ should be much larger than one. As a consequence, the expected recovery does not scale linearly with $\alpha L$. A high initial absorption is very important for the storage. As an example, a 80\% efficiency would typically require $\sqrt{\alpha L} \simeq 100$. This is a current experimental limitation \cite{phillips:023801}.
The memory time is also a common feature of EIT and SHBSL. The signal is finally mapped into the Raman coherence whose lifetime is limiting the storage. The protocols being designed for different systems, atomic vapors or REIC, these lifetimes can be very different.
Another common characteristic is that they both correspond to an adiabatic following of the atomic variable. For the SHBSL, the off-resonant optical coherences are following the field \cite{grischkowsky1973afa}. For the EIT, the Raman coherences are directly involved in the adiabatic process.
On the contrary, the nature of the absorption profile is very different. EIT is well suited for homogeneous system even if in practice it may be implemented with an effective reduced inhomogeneous broadening prepared by spectral tailoring \cite{longdell2003sls}. The SHBSL requires an inhomogeneous linewidth where a narrow hole is burnt.
The techniques to generate the TW are also very different. On on side, the transparency is produced by applying a coupling field that coherently couples the excited state to the second ground state. On the other side, the hole is prepared by incoherent optical pumping. In the latter case, the TW can persist for a long time. This is an interesting advantage since the preparation pulses are off during the storage sequence. Even if the Raman read-out pulse can be a source of detection noise, it is temporally separated from the recovered signal.
The protocols EIT and SHBSL have some common feature but are also very different by nature. This comparison should help to understand both. In practice, they have been conceived for different systems whose properties are well matching the advantages of each protocol.

\section{Atomic Frequency Comb}\label{AFC}
The storage mechanism of the AFC protocol is different by nature even if the techniques to prepare the sample and manipulate the coherences are the same. For the AFC, the interaction bandwidth is defined by a periodic distribution of very absorbing peaks. The medium exhibits very absorbing and fully transparent windows alternatively. As we will see later, the first observation that we can make is the exponential decay of the incoming field as it propagates. The primary absorption of the signal is also associated with a direct excitation of the atomic dipoles. In that sense, the mapping of the field in the medium is intuitive since the absorption profile covers the incoming signal bandwidth. The price will be paid at the retrieval stage because the recovered signal may be re-absorbed then reducing the whole protocol efficiency. The AFC technique has been designed to solve this apparent contradiction. 

The protocol first demands to tailor the inhomogeneous broadening. By optical pumping, a periodic absorbing structure is prepared. The comb peaks should be ideally narrow and very absorbing. We will first develop a formalism describing the storage in a periodically structured medium. This allows to connect the AFC and the well-known three-pulse photon echo technique. We will then show the results obtained in Tm:YAG. We will finally present the principle of the backward configuration leading to a perfect efficiency.

\subsection{Propagation through a spectrally periodic medium}\label{AFC_fw}
We assume here that the medium has a periodic spectral distribution $\displaystyle g\left(\Delta\right)$. We then consider the propagation of a weak signal through this structure. The periodic distribution generates an echo in the temporal domain. Our goal in this section is to calculate the echo efficiency. Our formalism is general and allows the connection between the AFC and the standard three-pulse photon echo.

The formation of an echo can be interpreted in the temporal domain as a rephasing of the coherences. The dipoles are initially excited by the incoming field. They rephase during their free evolution each time the phases equals an integer number of $2\pi$. If the spectral period is $1/T$, one expect a series of successive echos at different times $pT$ ($p\geq0$, then $p=0$ is the transmitted signal, $p=1$ is the first echo and so on). Alternatively in the frequency domain, if the spectral distribution is periodic, it acts as a grating whose successive orders correspond to the echos. We decompose this distribution in a Fourier series $\displaystyle g\left(\Delta\right)= \sum_n g_n e^{-i n \Delta T}$. If the structure is uniform over signal bandwidth, we expect the echo to have the same temporal shape as the incoming signal $\Omega(z=0,t)$. The respective amplitudes $a_p$ are then only z-dependent: $\displaystyle \Omega(z,t)=\sum_{p\geq0} a_p\left(z\right) \Omega\left(0,t-pT\right)$. Or equivalently in the Fourier domain: $\displaystyle \widetilde{\Omega}\left(z,\omega\right)= \widetilde{\Omega}\left(0,\omega\right) \sum_{p\geq0} a_p\left(z\right) e^{-i p \omega T}$. The coefficient $a_p$ is the diffraction efficiency of the p-th order. It can be calculated from the Fourier series of $g\left(\Delta\right)$. We use Bloch-Maxwell equations (eqs. \ref{MB_M} and \ref{MB_B}) in the frequency domain. We have dropped the usual term $\partial_t\Omega(z,t)$ because realistically the spatial extension of the pulse is in that case much longer than the length of our crystal.

\begin{equation}\label{MB_F}
\begin{array}{ll}
\partial_z\widetilde{\Omega}\left(z,\omega\right)=
-\displaystyle\frac{i \alpha}{2\pi}\int_\Delta g\left(\Delta\right) \widetilde{\mathcal{P}}\left(\Delta;z,\omega\right) \\[0.4cm]
\widetilde{\mathcal{P}}\left(\Delta;z,\omega\right)=-\displaystyle\frac{\widetilde{\Omega}\left(z,\omega\right)}{\omega+\Delta-i\gamma}
\end{array}
\end{equation}
The propagation equation can now be written as a Fourier expansion. In the forward direction:
$$
\partial_z\widetilde{\Omega}\left(z\right)=\displaystyle\frac{i \alpha}{2\pi} \widetilde{\Omega}\left(0,\omega\right) \sum_{p\geq0,n} g_n a_p\left(z\right) e^{i \left(n-p\right)  \omega T }  \int_\Delta \frac{e^{ -i n  \Delta' T }}{\Delta'-i\gamma}
$$
We here recognize an integral representation of the Heaviside step function $Y$ in the no-damping limit $\gamma {\rightarrow} 0$ (with $Y\left(0\right)=1/2$). The  coefficients are recursively given by the relation:

\begin{equation}\label{a_p}
\partial_z a_k\left(z\right) = -\alpha \sum_{p=0}^\infty a_p\left(z\right) g_{p-k} Y\left(k-p\right) 
\end{equation}

The diffraction order $k$ is imposed to be larger than $p$ for causality reason: the generation of the k-th echo depends only on the propagation of the previous ones ($0\leq p \leq k$). In this expression, the Heaviside function takes only two possible values, $1/2$ if $p=k$ and $1$ otherwise.

\begin{equation}\label{a0a1}
\begin{array}{ll}
\partial_z a_0\left(z\right) =\displaystyle -\frac{\alpha}{2} g_0  a_0\left(z\right) \\[0.4cm]
\partial_z a_1\left(z\right) =\displaystyle -\frac{\alpha}{2} \left[g_0  a_1\left(z\right) + 2 g_{-1}  a_0\left(z\right) \right]
\end{array}
\end{equation}

We define the efficiency $\eta$ as the intensity of the first echo \textit{i.e.} $|a_1\left(L\right)|^2$ where $L$ is the length of the crystal. It can be calculated by integrating the equations \ref{a0a1} with the boundary conditions $a_0\left(0\right)=1$ matching the incoming field intensity and $a_1\left(0\right)=0$:
\begin{equation}\label{etag0g1}
\eta\left(\alpha L\right)  =\displaystyle \left(g_{-1} \alpha L \right)^2  e^{-g_{0} \alpha L }
\end{equation}
where $L$ is the length of the crystal

This formalism allows us to describe a wide range of situations since we only assume the spectral periodicity of the medium and consider the propagation in the forward direction. We can for example calculate the efficiency of the three-pulse photon echo (3PE). In that case, a signal is diffracted by a modulation of the population: $\displaystyle g\left(\Delta\right)=\left[1+\cos\left(\Delta T\right)\right]/2$. We can then calculate the efficiency from $g_{0}$ and $g_{-1}$:

\begin{equation}\label{eta3PE}
\eta_{3PE}\left(\alpha L\right)  =\displaystyle \left( \alpha L/4 \right)^2  e^{ -\alpha L/2 }
\end{equation}

This function reaches a maximum of 13.4\% for $\alpha L=4$. This shows the trade-off between the absorption of the signal itself and the re-absorption of the emitted echo. This is in good agreement with what have been observed experimentally in non-inverted mediums \cite{Zafarullah2007158}.

In the prospect of quantum light storage, the necessary fidelity demands to consider a new regime. A complete recovery will first require for the signal to be fully absorbed. In the AFC protocol, the peaks should be very absorbing and narrow as compared to the comb spacing. This situation allows a significant gain in efficiency. Various shapes are possible for the spectral comb. This is a degree of control that can be used to optimize the efficiency. So far, we only consider a structure composed of well-separated lorentzian peaks $\displaystyle g\left(\Delta\right)= \sum_{l} \frac{1}{1+\left(\Delta-2\pi l/T\right)^2/\Gamma^2}$. The width $\Gamma$ is supposed to be smaller than the comb period l/T. The coefficients $g_{0}$ et $g_{-1}$ are then simply given by the Fourier transform of the Lorentzian: $g_{0}=\Gamma T/2$ and $g_{-1}=\Gamma T/\left(2 e^{\Gamma T}\right)$. If we define the comb finesse has $F=\pi/\left(\Gamma T\right)$, the efficiency in the forward direction is 
\begin{equation}\label{etaAFC}
\eta_{AFC}\left(\alpha L,F\right)  =\displaystyle \left( \frac{\pi \alpha L}{2F} \right)^2  \displaystyle e^{ -\frac{\pi \alpha L}{2F}-\frac{2\pi}{F} }
\end{equation}

When the optical thickness $\alpha L$ is high, the width of the peak should be properly chosen to optimize the efficiency. The optimal finesse is $F=\pi \left(1+d/4\right)$ leading to the maximum efficiency:

\begin{equation}\label{effopt}
\eta_{AFC}^{opt}\left(\alpha L\right)=4e^{-2} \left[\frac{\alpha L}{4+\alpha L}\right]^2
\end{equation}

\begin{figure}[pth]
\includegraphics[width=7.5cm]{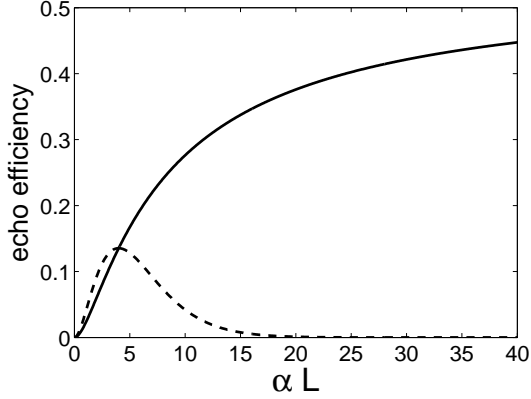}\caption{Efficiency comparison of the 3PE (dashed line) and the AFC (solid line) in the forward direction. If the comb finesse is properly adjusted, the efficiency of the AFC increases and reaches its maximum value of 54\%.}
\label{fig:Fig_AFCvs3PE}%
\end{figure}

Even if the efficiency is limited to $4e^{-2}$=54\% in the forward direction, it corresponds to a significant improvement as compared to the 3PE(see fig. \ref{fig:Fig_AFCvs3PE}).

Our analysis shows that an atomic frequency comb composed of narrow but very absorbing peaks is particularly efficient in terms of echo generation. Our first treatment is limited to the propagation in the forward direction. This precisely corresponds to the experimental situation that has been implemented in Tm:YAG as we will see now. The AFC protocol actually involves a backward retrieval. In that case, the efficiency is not limited to 54\% and goes to 100\%. It makes the protocol ideally perfect for quantum memory application. The procedure will be described in the last section \ref{AFC_complete}

\subsection{AFC in Tm:YAG}\label{AFC_TmYAG}
For the realization of the AFC in the forward direction, we use the same crystal as described in section \ref{SHBexp}. Since we are interested in the same properties \textit{i.e.} narrow homogeneous width and long hole lifetime, this choice is appropriate. We then should be able to tailor spectrally fine and optically deep absorbing feature defining the atomic comb. The experimental setup is the same as in section \ref{SHBexp} except for the crystal orientation in the magnetic field. In the present case, the polarization and the magnetic field are all oriented along the [001] axis. Even if the branching ratio is lower that the optimized configuration, it allows to equivalently excite all substitution sites \cite{louchet:035131}. We apply a 210G magnetic field. It splits the ground and the excited into a nuclear spin doublet by $\Delta_g=6 $ MHZ and $\Delta_e=1.3$ MHZ \cite{de2006experimental}.

The AFC preparation is performed by applying a sequence of weak pulse pairs (duration $\sim 300 \mathrm{ns}$ separated by $T=1.5 \mathrm{\mu s}$ and then followed by a $100 \mathrm{\mu s}$ dead time before the next pair). A pulse pair is a simplest pattern with a periodic spectral structure. Even if the spectral distribution is also imposed by the optical pumping dynamics, it is sufficient to produce a comb like structure. A preparation train consists of 5000 pairs followed by a long waiting time ($50 \mathrm{ms}$) to ensure the complete decay of population from the excited state (lifetime $\sim 800 \mathrm{\mu s}$). Since a single frequency pumping produces side holes at $\pm\Delta_e$, an AFC preparation procedure generates side shifted combs. To get around this effect, we make the comb period $1/T$ coincide with the excited state level splitting ($\Delta_e=2/T$ in our case).

\begin{figure}[pth]
\includegraphics[width=7cm]{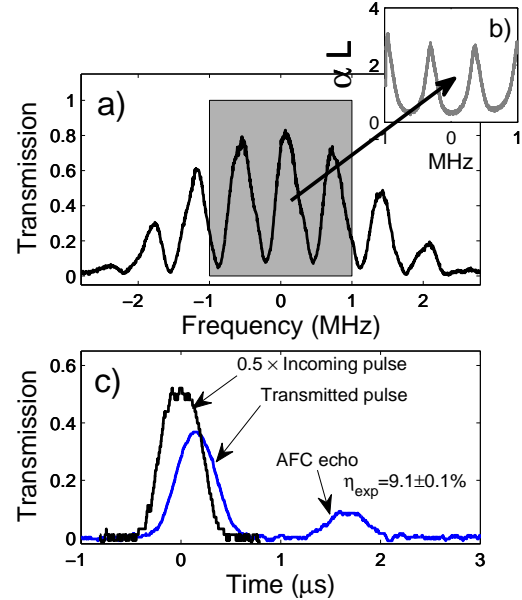}\caption{a) Transmission spectrum after the preparation procedure. b) Optical thickness for the central part of the spectral distribution. c) A weak pulse is sent (black curve). We observe an AFC echo (red curve).}
\label{fig:FigAFC_JLumin}%
\end{figure}

To probe the spectral distribution, we use a second beam whose frequency is swept by an acousto-optic modulator (AOM). We obtain the transmission spectrum (see fig. \ref{fig:FigAFC_JLumin} a.). The optical thickness is then simply $g\left(\Delta\right) \alpha L$ (see fig. \ref{fig:FigAFC_JLumin} b.). From this measurement, we can extract the coefficients $g_{0}$ and $g_{-1}$ by a numerical Fourier transform. We can then predict the expected diffraction efficiency of the comb $\eta=7.8\%$ in that case.

To verify the storage capability of the medium and measure the experimental efficiency, we  send a very weak signal pulse. Its duration is chosen to be slightly longer than the preparation pulses (FWHM $\sim 450 \mathrm{ns}$). It ensures a good spectral overlap with the comb. Part of the pulse is transmitted (see fig. \ref{fig:FigAFC_JLumin} c.) with an amplitude $|a_0\left(L\right)|^2$. We also observe the formation of an echo at time $T$. To obtain a reference pulse, we bleach the sample by applying a strong hole-burning beam (a comparable method as been used in section \ref{SHBexp}). In that case, the weak probe pulse is fully transmitted (black curve in fig. \ref{fig:FigAFC_JLumin} c.). By comparing this reference with the AFC echo, we measure the efficiency $\eta_{exp}= 9.1 \pm 0.1 $\%. This is in reasonnable agreement with the expected efficiency $\eta=7.8\%$ deduced from the spectral distribution. This slight difference can be due to the difficulty of measuring large optical thicknesses with accuracy \cite{chaneliere2009efficient}.

\subsection{Backward retrieval of the AFC}\label{AFC_complete}
The spectral preparation is a crucial preliminary step for the realization of the protocol. It allows us to observe good efficiencies in the forward direction. To be faithful, the complete procedure demands a backward retrieval. This point is relatively intuitive since the retrieved signal does not propagate through an absorbing medium. Very general time-reversal arguments can be invoked to understand this requirement \cite{moiseev-2008}.

\begin{figure}[pth]
\includegraphics[width=7cm]{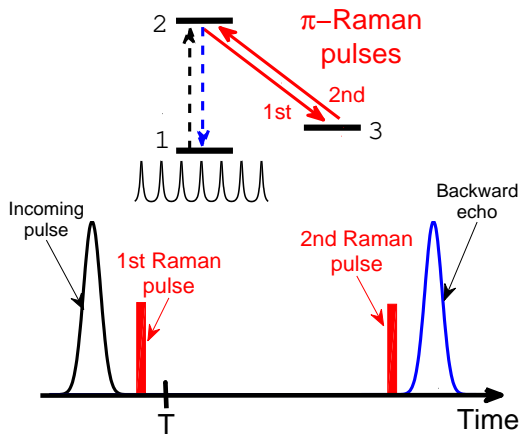}\caption{After periodic spectral preparation, a weak pulse is sent into the AFC medium. Without Raman pulses, an AFC echo would still appear at time T. This situation has been studied experimentally in section \ref{AFC_TmYAG}. The two counter-propagating Raman conversion pulses allow to store first the information in the long-lived ground state coherences and to then trigger the retrieval in the backward direction.}
\label{fig:AFCprotocol}%
\end{figure}

In practice, the retrieval direction can be flipped from forward to backward thanks to a pair of counter-propagating Raman pulses (see fig. \ref{fig:AFCprotocol}). This manipulation imprints a spatial $2kz$ phase factor modifying the phase matching condition and then changing the retrieval direction. If the signal is mapped into the optical coherences at time $t=0$, without Raman conversion, one would expect a retrieval at $t=T$. The generation of an echo is interpreted as a coherence rephasing. This evolution can be interrupted by a first Raman transfer ($\pi$-pulse). This has the advantage to freeze the atomic evolution. The storage time is then only limited by the lifetime of the Raman coherence. To recover the signal a second counter-propagating Raman $\pi$-pulse is applied. The evolution of the optical coherences is then resumed and gives rise to the retrieval in the backward direction. The Raman transfer is an important step since it permits i) a long memory time. It is now only given by the coherence lifetime in the ground state. ii) It modifies the phase matching condition and allows the backward retrieval. iii) It finally permits an on-demand signal recovery. This last point is important for a quantum repeater whose core is a quantum memory.

As in section \ref{SHBexp}, Tm:YAG may not be appropriate for the implementation of the backward retrieval. For the forward configuration (section \ref{AFC_TmYAG}), we use the Zeeman state as population shelving state to prepare the atomic comb. We then cannot apply Raman pulses on this non-empty level otherwise this would bring some population in the excited state and would have deleterious consequences for a quantum memory \cite{ruggiero:053851}. Nevertheless Tm:YAG has some original advantageous features as mentioned in the introduction. It is then a unique test-bed to conveniently and accurately study the figures of merit of the protocol \textit{i.e.} the efficiency, the bandwidth, the multi-mode capacity \cite{nunn} and the interplay between them.

\section{Conclusion}

We have studied two quantum storage protocols namely the Spectral hole burning for Stopping Light and the Atomic Frequency Comb protocol. They are both well suited to REIC since they fully use their specific properties. Even if the storage mechanisms are fundamentally different, they involve the same techniques to manipulate the medium. For both, the preparation is based on hole-burning techniques which permits spectral tailoring of the absorption line. They also employ a Raman transfer to convert the optical excitation into long-lived ground state coherences. For the theoretical description, we apply the same formalism.

The critical stages of the protocols have been implemented and studied experimentally in a Tm:YAG crystal. Different routes are currently under investigation in our group to improve the feature of the storage step and pave the way toward quantum memory application.




\bibliographystyle{elsarticle-num}

\end{document}